\documentclass[lettersize,journal]{IEEEtran}
\usepackage{cite}
\usepackage{amsmath,amssymb,amsfonts}
\usepackage{mathtools}
\usepackage{algorithm}
\usepackage[noend]{algpseudocode}
\usepackage{graphicx}
\usepackage{textcomp}
\usepackage{xcolor}
\usepackage{siunitx}
\usepackage{hyperref}
\usepackage{cleveref}
\usepackage{xspace}
\usepackage{caption,subcaption}
\usepackage{multirow}
\usepackage{makecell}
\usepackage{color, colortbl}
\usepackage{csquotes}
\usepackage{booktabs}
\usepackage[abbreviations]{foreign}
\def\BibTeX{{\rm B\kern-.05em{\sc i\kern-.025em b}\kern-.08em
    T\kern-.1667em\lower.7ex\hbox{E}\kern-.125emX}}

\algnewcommand{\LeftComment}[1]{\State {\color{violet} \slash* #1 *\slash}}
\algrenewcommand{\algorithmiccomment}[1]{{\color{violet}\hfill$\triangleright$ #1}}

\newif\iftodos
\todostrue %

\definecolor{darkgreen}{rgb}{0.01, 0.75, 0.24}
\definecolor{darkblue}{rgb}{0.01, 0.01, 0.54}

\newcommand{\SystemName}{\textsc{Plexus}}

\usepackage{acronym}
\acrodef{DL}{decentralized learning}
\acrodef{GL}{gossip learning}
\acrodef{FL}{federated learning}
\acrodef{D-PSGD}{decentralized parallel stochastic gradient descent}
\acrodef{SGD}{stochastic gradient descent}

\newcommand{\femnist}{FEMNIST\xspace}
\newcommand{\celeba}{CelebA\xspace}

\newcommand{\cifar}{CIFAR-10\xspace}
\newcommand{\movielens}{MovieLens\xspace}
\newcommand{\leaf}{LEAF\xspace}

\newcommand{\dpsgd}{{\xspace}\ac{D-PSGD}\xspace}

\newcommand{\view}{local view\xspace}
\newcommand{\views}{local views\xspace}

\begin{document}

\algblockdefx[Request]{Request}{EndRequest}{\textbf{request}~}{\textbf{end}}
\makeatletter
\ifthenelse{\equal{\ALG@noend}{t}}{\algtext*{EndRequest}}{}
\makeatother
\algblockdefx[Upon]{Upon}{EndUpon}{\textbf{upon}~}{\textbf{end}}
\makeatletter
\ifthenelse{\equal{\ALG@noend}{t}}{\algtext*{EndUpon}}{}
\makeatother
\newcommand{\send}{\textbf{send to}~}
\algblockdefx[AsyncFun]{AsyncFun}{EndAsyncFun}{\textbf{async function}~}{\textbf{end}}

\title{Decentralized Learning Made Practical with Client Sampling}

\author{
\IEEEauthorblockN{Martijn de Vos\IEEEauthorrefmark{1}\IEEEauthorrefmark{3}, Akash Dhasade\IEEEauthorrefmark{1}, Anne-Marie Kermarrec\IEEEauthorrefmark{1}, Erick Lavoie\IEEEauthorrefmark{2}, Johan Pouwelse\IEEEauthorrefmark{3}, Rishi Sharma\IEEEauthorrefmark{1}}
\IEEEauthorblockA{\IEEEauthorrefmark{1}EPFL, Switzerland\\}
\IEEEauthorblockA{\IEEEauthorrefmark{2}University of Basel, Switzerland\\}
\IEEEauthorblockA{\IEEEauthorrefmark{3}Delft University of Technology, The Netherlands}
}

\maketitle

\begin{abstract}

Decentralized learning (DL) leverages edge devices for collaborative model training while avoiding coordination by a central server.
Due to privacy concerns, DL has become an attractive alternative to centralized learning schemes since training data never leaves the device.
In a round of DL, all nodes participate in model training and exchange their model with some other nodes.
Performing DL in large-scale heterogeneous networks results in high communication costs and prolonged round durations due to slow nodes, effectively inflating the total training time.
Furthermore, current DL algorithms also assume all nodes are available for training and aggregation at all times, diminishing the practicality of DL.
This paper presents \SystemName{}, an efficient, scalable, and practical DL system.
\SystemName{} (1) avoids network-wide participation by introducing a decentralized peer sampler that selects small subsets of available nodes that train the model each round and, (2) aggregates the trained models produced by nodes every round.
\SystemName{} is designed to handle joining and leaving nodes (churn).
We extensively evaluate \SystemName{} by incorporating realistic traces for compute speed, pairwise latency, network capacity, and availability of edge devices in our experiments.
Our experiments on four common learning tasks empirically show that \SystemName{} reduces time-to-accuracy by 1.2-8.3$\times$, communication volume by 2.4-15.3$\times$ and training resources needed for convergence by 6.4-370$\times$ compared to baseline DL algorithms.
\end{abstract}


\section{Introduction}
\label{sec:introduction}

\Ac{DL} systems empower edge devices (referred to as \emph{nodes} in this paper) to collaboratively train a machine learning model without sharing their raw data.
In \ac{DL} systems, each node maintains a local model.
Every round, a node trains its local model on its own data, sends the trained model to a subset of other nodes, and aggregates incoming models with its local model.
A communication topology dictates how nodes exchange their models~\cite{lian2017can,pmlr-v119-koloskova20a}.
\ac{DL} offers several advantages over centralized training methods, including
alleviation of the communication and computational load on centralized servers and more efficient training on large-scale datasets~\cite{ormandi2013gossip,lian2017can,pmlr-v80-lian18a,yu2019linear}.
Furthermore, the private data of the participating nodes in \ac{DL} does not leave their devices at any point during the training process, therefore enhancing privacy of user data.
Finally, \ac{DL} can significantly reduce operational costs compared to training within data centers since no dedicated infrastructure is needed.
\ac{DL} is increasingly explored in industrial scenarios, for example, in healthcare~\cite{kasyap2021privacy,tedeschini2022decentralized,lu2020decentralized} or Internet-of-Things settings~\cite{lian2022decentralized,gerz2022comparative}.

While the convergence and performance of \ac{DL} systems have been extensively studied from a theoretical perspective~\cite{pmlr-v119-koloskova20a,pmlr-v97-koloskova19a,pmlr-v139-kong21a,dandi2022data,liu2022decentralized}, their deployment in edge settings with a potentially large number of devices remains scarce~\cite{bellavista2021decentralised}.
In edge scenarios at scale, the participating devices can hold heterogeneous data~\cite{pmlr-v54-mcmahan17a}.
Moreover, the devices exhibit substantial disparities in their computational capabilities~\cite{lai2022fedscale}.
These two forms of heterogeneity impact the convergence and performance of \ac{DL} systems~\cite{hsieh2020non,chen2022fairness}.

Furthermore, compounding the challenge, the edge devices also demonstrate churn, implying that these devices can join or leave the network at any given moment~\cite{lai2022fedscale,abdelmoniem2023refl}.
For instance, our analysis of commonly used real-world availability traces~\cite{lai2022fedscale} revealed that only 8.8\% of the devices are online at any given time in the best case. 
Current \ac{DL} systems are not designed to handle churn.
This oversight is not surprising as DL systems originate from data center environments~\cite{lian2017can} where nodes are more reliable compared to those in edge settings.
It remains an open question how to design a \ac{DL} system that operates efficiently in heterogeneous and large-scale edge environments with nodes joining and leaving~\cite{bellavista2021decentralised}.

In this paper, we present the design, implementation, and evaluation of \SystemName{}, a novel practical system for decentralized learning. 
The core of \SystemName{} lies in its \textit{decentralized peer sampler} and \textit{sample-wide aggregation} mechanism capable of handling churn.
The peer sampler enables nodes to independently determine a subset of online nodes, or a \emph{sample}, that is in charge of the training process for a given round.
Since not all available nodes are expected to train each round, the peer sampler significantly reduces the resource consumption of the training process compared to traditional \ac{DL} systems.
The sample changes every round, therefore, evenly balancing the training load over nodes and providing online nodes with equal opportunity to contribute to model training.
Following local model training, the sample-wide aggregation scheme selects a single \textit{aggregator} from within the sample to aggregate all trained models generated in each round.
This aggregation process accelerates model training and reduces time-to-accuracy, surpassing traditional DL systems that rely on local model aggregation within a node's immediate neighborhood.

We evaluate \SystemName{} using real-world mobile phone traces of pairwise latencies, bandwidth capacities, computation speed, and availability~\cite{lai2022fedscale}.
Our evaluation covers four learning tasks in varying network sizes, up to \num{1000} nodes.
We compare the \SystemName{} performance with Gossip Learning~\cite{hegedHus2019gossip} and D-PSGD~\cite{lian2017can}, two state-of-the-art \ac{DL} algorithms.
Our experimental results show that \SystemName{} reduces time-to-accuracy by 1.2-8.3$\times$, communication volume by 2.4-15.3$\times$, and training resources consumed by 6.4-370$\times$.
We also experiment with different sample sizes (\ie, the number of nodes that train each round), demonstrate that \SystemName{} incurs minimal communication overhead, and empirically establish that \SystemName{} rapidly disseminates and synchronizes availability information across nodes in the network to effectively handle churn.

In summary, this work makes the following contributions:

\begin{enumerate}
    \item We design \SystemName{}, a practical \ac{DL} system tailored to large-scale and real-world network of nodes (\Cref{section:system}).
    \SystemName{} incorporates a decentralized peer sampler to select a small subset of nodes that train the model each round, significantly reducing training resources required to converge compared to existing approaches.
    \SystemName{} then leverages a single aggregator node per round for sample-wide model aggregation, accelerating the training process.
    To handle churn, nodes announce their network membership status (joined or left) upon change to other nodes; membership information of nodes is piggy-backed on model transfer messages.
    Our system operates without any centralized or network-wide coordination.
    
    \item We conduct an extensive evaluation of \SystemName{} using real-world mobile phone traces at scale, comparing it with prominent baseline \ac{DL} algorithms (\Cref{sec:eval}). Our results demonstrate that \SystemName{} significantly enhances performance across four common learning tasks in terms of time-to-accuracy, communication volume, and resource requirements.
    
    \item We provide an open-source implementation of \SystemName{} on GitHub, fostering reproducibility and encouraging future research of \ac{DL} in heterogeneous edge settings.
\end{enumerate}

\section{Towards Practical Decentralized Learning}
\label{sec:motivation}
\Ac{D-PSGD} is the most common algorithm to realize \ac{DL} and operates in a series of synchronous training rounds~\cite{lian2017can,zhu2022topology}.
During each round, nodes perform local model training, send their trained model to their \textit{immediate} neighbors and aggregate received models with its local model.
A communication topology, set before the start of the learning process, dictates model exchange.
Nodes await the reception of models from all their neighbors before advancing to the next round.
Despite the frequent use of \dpsgd for \ac{DL}, we formulate two notable shortcomings of \dpsgd that make its deployment in large-scale edge settings challenging.

\begin{figure}[t]
	\centering
	\begin{subfigure}{.5\columnwidth}
		\centering
		\includegraphics[width=\linewidth]{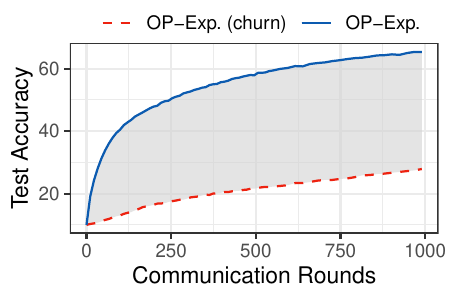}
        \captionsetup{width=.9\linewidth}
        \caption{Churn can significantly impact convergence.}
		\label{fig:motivation_churn}
	\end{subfigure}%
	\begin{subfigure}{.5\columnwidth}
		\centering
		\includegraphics[width=\columnwidth]{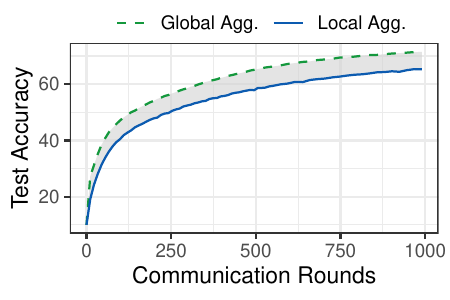}
        \captionsetup{width=.9\linewidth}
        \caption{Global aggregation improves convergence.}
		\label{fig:motivation_aggregation}
	\end{subfigure}%
	\caption{The effect of churn (left) and the convergence of local and global aggregation (right) in \dpsgd.}
    \vspace{-5pt}
	\label{fig:motivation}
    \vspace{-10pt}
\end{figure}

\textbf{Dealing with Churn.}
In cross-device learning settings, devices typically have intermittent availability for training~\cite{bonawitz2019towards,kairouz2021advances,abdelmoniem2023refl}.
\dpsgd was not initially crafted with churn mitigation in mind and offline nodes can therefore significantly deteriorate the convergence of \dpsgd. 
To illustrate the impact of churn on \dpsgd, we chart in~\Cref{fig:motivation_churn} the test accuracy for \dpsgd both in the presence and absence of churn on the \cifar dataset under an Independent and Identically Distributed (IID) data distribution.
We implement \dpsgd with a one-peer exponential graph topology (OP-Exp.) which is considered a state-of-the-art \ac{DL} algorithm~\cite{ying2021exponential}.
This topology provides fast propagation of model updates through the network as the maximum distance between any pair of nodes is $ O(log\ n) $.
With this topology, each node receives and sends exactly one model every round.
A peer is connected to $ log(n) $ neighbors ($ n $ is the total network size) and cycles through them round-robin.
To evaluate the setting with churn, we assume that 8.8\% of nodes are online in any given round; this number is derived from the \emph{best-case} scenario in a real-world trace containing mobile device availability~\cite{yang2021characterizing}.
\Cref{fig:motivation_churn} highlights that node churn significantly hurts convergence.
In the experiment with churn, at any given time, only a fraction (8.8\%) of nodes are online for training and aggregation, which potentially may not be immediate neighbors, leading to poor convergence.
The shaded area in \Cref{fig:motivation_churn} shows the room for improvement if we design our system to handle node churn effectively.

\textbf{Local vs. Global Aggregation.}
\dpsgd aggregates models amongst neighborhoods, i.e., local aggregation~\cite{lian2017can}.
Local aggregation leaves \emph{residual variance} between local models, biasing gradient computations and slowing down model convergence compared to when doing global aggregation~\cite{pmlr-v139-kong21a,bellet2021d}.
The optimal convergence speed, however, is achieved when aggregating all produced models each round, i.e., global aggregation.
However, this becomes costly in terms of communication volume as an all-to-all model exchange is needed.
We illustrate this effect in~\Cref{fig:motivation_aggregation}, showing the evolution of test accuracy for \dpsgd with global aggregation (\ie with a fully-connected topology) and local aggregation (using a one-peer exponential graph topology).
We observe that global aggregation is beneficial for convergence by nullifying residual variance across nodes in the network.
This experiment is performed in a no-churn scenario and the room for improvement is shown in grey in \Cref{fig:motivation_aggregation}.

In summary, the design of \SystemName{}, explained in the next section, is based on the following two key insights:
\begin{enumerate}
    \item Not all nodes can or have to participate in training every round.
    \item Performing global aggregation at the end of each round is highly beneficial for model convergence.
\end{enumerate}

\section{Design of \SystemName{}} %
\label{section:system}

We first describe our system model and assumptions in Section~\ref{section:system-model}, provide a conceptual overview of the algorithm in Section~\ref{section:overview}, and then present the components of \SystemName{} in the remaining sections.%

\subsection{System Model and Assumptions}
\label{section:system-model}

We consider a peer-to-peer network of $n$ nodes that collaboratively train a global machine learning model $ \theta $. Each participating node has access to a local dataset which never leaves the participants' device.  Only the model parameters are exchanged between participating nodes. We assume that each node knows the specifications of the ML model being trained, the learning hyperparameters, and the settings specific to \SystemName{}.
These can be exchanged before training starts.

\textbf{Network and failure model.} 
In contrast to a data center setting, model training with \SystemName{} proceeds in a decentralized environment and relies on the cooperation of nodes with intermittent availabilities and varying resource capacities.
We assume that each node has a unique identifier (\eg, a public key) and assume that nodes are connected through a fully connected overlay network (\textit{i.e.}, all nodes can communicate with each other).
Nodes may join or leave the network at any time.
Though the computational capacities of participating nodes may vary, we assume that each node's computational resources are sufficient to reliably participate in learning.
We assume that the model being trained fits into the memory of individual nodes.
Also, aggregators in \SystemName{} should have sufficient memory or disk space to store and aggregate the trained models produced by other nodes.
Finally, we assume a partially synchronous network model in which the delivery time of network messages is periodically bounded~\cite{dwork1988consensus}. 
As such, we can detect unresponsive nodes with timeouts during the periods of synchrony and otherwise wait until the network becomes synchronous again.

In designing and evaluating \SystemName{}, our primary focus is on system scalability and efficiency under churn.
While we remark that nodes might act malicious during the training process, \eg, by data poisoning or adversarial attacks, this introduces a separate layer of complexity and we consciously leave out these considerations.
However, we acknowledge research in privacy-preserving ML, many of which we believe could be integrated or adapted into \SystemName{}~\cite{bonawitz2017practical,mothukuri2021survey}.

\begin{figure}[t]
    \includegraphics[width=\linewidth]{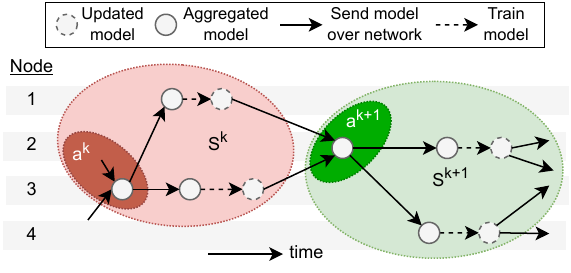}
    \caption{Overview of round $ k $ and $ k+1 $ in \SystemName{}, including 4 total nodes ($n = 4$)  and a sample size of 2 ($s = 2 $). 
    Nodes 1 and 3 are in sample $ S^k $ and first train the aggregated model they received on their local data.
    They then send their updated model to the aggregator $ a^{k+1} $ in sample $ S^{k+1} $. 
    When the aggregator receives all updated models, it aggregates the incoming models and forwards them to the participants in the next sample $S^{k+1}$.} 
    \label{fig:midas_overview}
    \vspace{-0.5cm}
\end{figure}

\subsection{\SystemName{} in a Nutshell} %
\label{section:overview} 

To save training resources, \SystemName{} avoids all nodes being involved in the training process every round and instead \textit{(i)} has a subset of online nodes (a \emph{sample}) train the model each round, and \textit{(ii)} refreshes samples each round.
We refer to nodes belonging to a sample as \emph{participants}.
Amongst the participants, one node, named the \emph{aggregator}, is responsible for model aggregation during that round.
This aggregator is selected to be the node with the highest bandwidth capacity as it has to handle incoming model transfers from all participants in a round.
In each round, participants are randomly sampled from all online nodes using a consistent hashing scheme.
This sampling mechanism is a key contribution of \SystemName{} and is outlined in~\Cref{subsec:deriving_samples}.

\Cref{fig:midas_overview} illustrates two rounds in \SystemName{}, both comprised of aggregation followed by training.
We show a training session with $ n = 4 $ nodes and sample size $s = 2$. 
We denote the set of nodes in the $k$-th sample as $ S^k $ and the aggregator within $ S^k $ as $ a^k $. 
In a given round $k$, the aggregator $a^k$ is activated when the participants of the previous sample $S^{k-1}$ push the updated models to the aggregator.
Upon aggregating, the aggregator sends this model to all $s$ participants in $S^k$. 
The participants train the model with their local data and then send their updated model to the next aggregator $a^{k+1}$ as the process repeats.

While the above description should give an idea of the functioning of \SystemName{}, we identify the following three technical challenges towards building a practical \ac{DL} system:
\begin{enumerate}
    \item \textit{Nodes going offline during training or aggregation:} How can \SystemName{} ensure system progression when nodes go offline during training or aggregation? We discuss this in \Cref{subsection:training-aggregating}.
	\item \textit{Decentralized sampling:} How can nodes derive consistent samples in a decentralized fashion, independently from other nodes? We discuss this in \Cref{subsec:deriving_samples}.
	\item \textit{Tolerating unavailability:} Since nodes can go offline, how can \SystemName{} avoid selecting offline nodes during sampling? We discuss this in \Cref{subsection:joining-leaving}.
\end{enumerate}

\subsection{Training and Aggregating Models}
\label{subsection:training-aggregating}
Each node in \SystemName{} implements two types of tasks that can run concurrently: one for aggregation and one for training.
This is because a node may concurrently train in a sample $S^k$ and aggregate in a sample $S^{k+1}$.
We associate separate round numbers with these aggregation and training tasks, referred to as $k_{agg}$ and $k_{train}$, respectively. 
The design of \SystemName{} is based on a \textit{push} model, in which nodes in sample $S^k$ trigger the activation of nodes in sample $S^{k+1}$. 
This way, nodes do not have to continuously be aware of the current training round being worked on; they only have to start working when receiving a trained or aggregated model. 

\textbf{Training.} We first describe the training procedure carried out by node $ i $.
A node starts the training task when it receives a \texttt{train} message, containing an aggregated model and a round number $ k $.
The round number $k_{train}$ is first used to interrupt a pending training task if the network has gotten ahead: training is interrupted if the same task is triggered with a higher round number ($ k > k_{train} $).
Node $ i $ then starts the model training task for the requested round number if it is not already training in round $ k $.
Once $ i $ completes its training, it sends an \textsc{Aggregate} message, containing the resulting model, to aggregator $ a $ in sample $ S^{k+1} $.
This aggregator is derived using our peer sampler (see \Cref{subsec:deriving_samples}).

To ensure the progression of learning, \SystemName{} needs to deal with the failure of aggregators.
To safeguard against the failure of aggregator $ a $, the nodes await an \textsc{Ack} message from $ a $ for a period of $\Delta t_{ack}$ time.
Once this timeout is triggered, the node retries by sampling another aggregator until it succeeds.

\textbf{Aggregation.}
An aggregator $ a $ starts the aggregation task when it is activated through an \texttt{aggregate} message.
$ a $ first checks whether the message is fresh ($k > k_{agg}$), ongoing ($k = k_{agg}$) or stale ($k < k_{agg}$).
A fresh message starts the aggregation task and $ a $ stores the received model.
Upon receiving more models for the ongoing round, $ a $ accumulates them.
Stale messages are responded to with an \textsc{Ack} message to the sender.
This prevents the stale sender from making further requests to other aggregators.

Participants might go offline before they finish sending their model to $ a $.
Therefore, $ a $ only requires the reception of some of the $ s $ models from participants to complete the aggregation.
We refer to the required fraction of models needed as the \emph{success fraction} $ sf $.
Specifically, $ a $ awaits $ \lfloor sf \times s \rfloor $ models before completing aggregation.
As an additional safeguard against offline participants, we also use an aggregation timeout $\Delta t_{agg}$.
This timeout starts when an aggregator receives the first trained model in a round, and when it expires, it finishes aggregation. 
Thus, learning progression does not stall even when $ a $ receives less than $ sf \times s $ models, and \SystemName{} continues as long as one reliable participant successfully pushes the trained model to the aggregator within $\Delta t_{agg}$.
The aggregator then informs the participants in the previous sample about the round completion by sending \textsc{Ack} messages.
Lastly, $ a $ aggregates the received models and sends a \texttt{train} message to the participants in the next sample as the process repeats.

\subsection{Deriving Samples}
\label{subsec:deriving_samples}

One of the main novelties of \SystemName{} is to decentralize the sampling procedure by having each participant in a sample compute the next sample independently. %
In order to achieve this, each node maintains a \emph{local view} of the network wherein the membership information of (all) other nodes is recorded.
The gist of \SystemName{}'s sampling procedure is to rely on a hash function parameterized by the round number and the node identifiers, stored by all nodes in their local view so that each node can independently compute the sample of nodes that are expected to be active during the training. 
\Cref{alg:get_sample} shows the \SystemName{} sampling procedure, which aims to obtain a sample of $s$ currently active nodes in the $k^{th}$ round.
First, a subset of \textit{candidates} that are considered online is retrieved to avoid unnecessarily waiting for known offline nodes.
We discuss how the online and offline status of nodes is synchronized in \Cref{subsection:joining-leaving}. 
Concatenating the node identifiers with round numbers randomizes the order of nodes every round. 
The resulting list is sorted in lexicographic order, which provides the order in which candidates are contacted.
As long as the list of candidates is mostly consistent between nodes, the order of contact and the resulting samples are mostly consistent.

A candidate may be marked as online in \views but could be offline.
When determining a sample, candidates are first contacted with a \texttt{ping} message, and only those that reply with a \texttt{pong} message before timeout $\Delta t_p$ are considered.
The first $s$ candidates are all contacted in parallel to lower latency. In the best and most common case, they all reply within $\Delta t_p$, and the procedure can return immediately.
Otherwise, the remaining candidates are contacted one by one until enough candidates have replied.
Our implementation associates a unique identifier with each \texttt{ping}-\texttt{pong} exchange for tracking messages.
This identifier is not included in \Cref{alg:get_sample} for presentation clarity.

Local views and therefore derived samples for a particular round may be temporarily inconsistent (\ie, partially non-overlapping) when \views are still being synchronized.
This results in participants send updated models to different aggregators or aggregators sending aggregated models to different participants.
However, we observed that the model variance in \SystemName{} introduced due to these inconsistencies is much lower than in \dpsgd where the variance between local models is high.
As \dpsgd has been proven to converge under this variance~\cite{lian2017can}, sporadic inconsistencies in \SystemName{} do not undermine the learning process.

\begin{algorithm}[t!]
	\caption{Sampling by node $i$ where $k$ denotes the round number and $s$ is the requested sample size.}
	\label{alg:get_sample}
	\begin{algorithmic}[1]
	    \State \textbf{Require}: Ping timeout $\Delta t_p$
	    \State
	    \Procedure{Sample}{$k,s$}

	         \LeftComment{\textsc{Actives}() are the online nodes in \views}
	        \State $H \leftarrow \textsc{sort}([ \textsc{hash}(j + k) ~\textbf{for}~ j ~\textbf{in}~ \textsc{Actives}() ])$

    	    \State $C \leftarrow [ j~\textbf{for}~h_j~\textbf{in}~H ]$ \Comment{Candidate identifiers}
                \State \textbf{return} the first $s$ in $C$ that answer a ping within $\Delta t_p$

	    \EndProcedure

	\end{algorithmic}
\end{algorithm}

\begin{algorithm}[t!]
	\caption{Aggregator selection from a sample by node $i$.}
	\label{alg:aggregator}
	\begin{algorithmic}[1]
		\Procedure{Aggregator}{$k,s$}
		\State $S^k \leftarrow \textsc{Sample}(k,s)$
		\State \Return $j \in S^k $ \textbf{such that} $j$ has the largest bandwidth among all $S^k$ nodes according to $B_i$
		\EndProcedure
	\end{algorithmic}
\end{algorithm}

We next present how participants determine an aggregator in \Cref{alg:aggregator}.
Internally, this method calls the sampling procedure from \Cref{alg:get_sample} (line 2).
The aggregator is a critical node for system progression and must handle the reception and transmission of at most $ s $ trained models during a round.
Since the aggregator has to handle this network load, we preferentially select the participant with the highest bandwidth capacity from the derived sample.\footnote{One approach to reduce communication burdens on aggregators is to use multiple aggregators in the same round. As this optimization requires additional coordination mechanisms, we leave this for further work.}
This biased aggregator selection optimizes the model transfer times and lower individual round durations.
We assume that bandwidth capacities of individual nodes are also included in the \views and gossiped with the model transfer messages.
We found that this decision was essential to the success of \SystemName{} as learning progress would slow down significantly if an aggregator with low bandwidth capacity is chosen, especially if the model size increases.
\SystemName{} randomizes uniformly the node sampling but we remark that a system designer can include other information in the \views, for example, memory or storage capacities or details on the available training hardware.
This information can be leveraged for a more guided selection of participants or aggregators~\cite{lai2021oort}.
Because samples are randomized uniformly in the current procedure, any online node in the network should be selected every $n/s$ times on average as a participant.

\subsection{Handling Joining and Leaving Nodes}
\label{subsection:joining-leaving}

\SystemName{} supports dynamic membership, \textit{i.e.}, participants can go online and offline, even when training or aggregating. 
To avoid sampling nodes that are offline, each node in \SystemName{} tracks the online or offline status of nodes in \views.
A \view $ V_i$, maintained by node $ i $, contains a dictionary $E_i$ that associates the most recent \texttt{joined} or \texttt{left} events to each node identifier.
Each \texttt{joined} and \texttt{left} event of a node $ i $ is associated with a local, persistent counter $c_i$.
The counter is incremented on every \texttt{join} or \texttt{left} event only by node $i$ itself, and therefore more recent events in the view can only originate from node $i$.
This counter is independent of the rounds of learning.
Therefore, an entry in $ E_i $ like \mbox{$<$2, (\texttt{join}, 3)$>$} indicates a \texttt{join} event with index 3 by the node with identifier 2.
Each node also maintains a dictionary $B_i$ in its local view, which stores the bandwidth information of other nodes in the network.
We note that this information is used for preferentially selecting the aggregator within a sample.
The tuple of dictionaries $(E_i, B_i)$ together form the complete \view $V_i$ of node $i$.

When joining for the first time or joining again after leaving the network, a node $i$ first increments its persistent counter $c_i$ and then updates its view with a new \textit{joined} event. 
It then advertises itself to a random set $P$ of nodes from the network with a \texttt{joined} message that includes its latest $c_i$ counter value.
Upon receiving a \texttt{joined} message from node $ i $ by node $ j $, $ j $ updates the corresponding view entry for $i$ with a \textit{joined} event, as long as the event is more recent than the last one recorded, \textit{i.e.}, the counter $c_i$ for that message is larger than the last one stored. 
The process for leaving is identical to that of joining, except that the event stored in the \view is \textit{left} instead of \textit{joined}.
Increasing the size of $ P $ helps to keep \views synchronized at the cost of additional communication.

Local views are exchanged between nodes by piggybacking them on the messages used to send models between participants (\textsc{train} and \textsc{aggregate} messages).
Since the models are typically several times larger than the views, the overhead of view synchronization is reasonable (also see~\Cref{sec:exp_overhead}). 
When node $ i $ receives the \view $V_j$ of another node $ j $, $ i $ merges $ V_j $ in its own \view by adopting the more recent events that may be contained $E_j$. 
Only the most recent event is kept in the \view because it is the only one necessary to determine whether a node is currently online or offline, \textit{i.e.}, its latest event being \textit{joined} or \textit{left}, respectively. %

In either cases of joining and leaving, if at least one node $j$ in $P$ is online and is selected as a participant for round $k$, node $i$ should become a candidate for round $k+1$.
This is because node $j$'s \view which will include $i$'s latest event, will propagate to the sample $k+1$.
Thereafter, each participant will be informed of node $i$'s availability status as the local views become more consistent.
In certain exceptional situations, it is possible that the \texttt{join} request of the node did not reach any reliable node, or the request messages were temporarily delayed.
Consequently, the network might remain oblivious to node $i$'s presence.
In those cases, a node may disseminate \texttt{join} messages to different nodes. 
By default, a node advertises its departure to at least one online node prior to going offline (\Cref{section:system-model}).
However, in some situations this is not possible, \eg, if a device runs out of battery or in the case of hardware failure.
\SystemName{} has safeguard mechanisms that handle delayed messages even from online nodes through the ping-pong timeouts ($\Delta t_p$), aggregator timeout ($\Delta t_{agg}$) and acknowledgment timeouts ($\Delta t_{ack}$) as discussed before.

\begin{table*}[t!]
    \caption{Summary of datasets used to evaluate \SystemName{} and \ac{DL} baselines.}
	\centering
	\begin{tabular}{c|c|c|c|c|c}
		\toprule
		\textsc{Dataset} & \textsc{Task} & \textsc{Nodes} & \textsc{Learning Parameters} & \textsc{Model} & \textsc{Model Size} \\ \hline
		CIFAR10~\cite{krizhevsky2009learning} & Image classification & \num{1000} & $ \eta = 0.002 $, momentum $ = 0.9 $ & CNN (LeNet~\cite{hsieh2020non}) & 346 KB \\ 
		CelebA~\cite{caldas2018leaf} & Image classification & 500 & $ \eta = 0.001 $ & CNN & 124 KB \\ 
		FEMNIST~\cite{caldas2018leaf} & Image classification & 355 & $ \eta = 0.004 $ & CNN & 6.7 MB \\ 
		MovieLens~\cite{grouplens:2021:movielens} & Recommendation & 610 & $ \eta = 0.2 $, embedding dim $= 20$ & Matrix Factorization & 827 KB \\ 
		\bottomrule
	\end{tabular}
	\label{table:experiment_datasets}
\end{table*}

\subsection{Setting \SystemName{} parameters}
\label{sec:setting_params}
We provide some guidelines to determine the \SystemName{} parameters.
The sample size $ s $ impacts the communication volume and resource usage during learning; increasing $ s $ also increases the load on the aggregator.
The ideal value of $ s $ highly depends on the network capacity of the aggregator in a deployment setting.
A higher success fraction $ sf $ likely prolongs round durations as an aggregator has to wait for more models.
We experiment with the values of $ s $ and $ sf $ in Section~\ref{sec:eval}.
To avoid incorrectly flagging a node as offline, one should set the ping timeout $\Delta t_p$ to the expected upper bound on the Round Trip Time of the underlying communication network.
To ensure all nodes have a chance to contribute to model training, the aggregation timeout $ \Delta t_{agg} $ should be set such that all participants in a round have sufficient time to complete both training and sending of their model to the aggregator.
Finally, $ \Delta t_{ack} $ should be greater than $ \Delta t_{agg} $ to prevent a participant from sending a model to another aggregator while the other aggregator is still waiting for models.

\subsection{Notes on \SystemName{} Convergence}
From a learning perspective, \SystemName{} follows a client-server paradigm analogous to \ac{FL}, thus the convergence proofs for \ac{FL} offer theoretical grounding for the model convergence of our approach~\cite{li2019convergence}.
However, it is imperative to note that, from a system design viewpoint, \SystemName{} diverges significantly from traditional \ac{FL} systems.
Unlike FL's inherent centralized orchestration, our design introduces a decentralized peer sampler and chooses a unique aggregator per round, eliminating the need for centralized coordination while dealing with node unavailability and network churn.

\section{Experimental Evaluation}
\label{sec:eval}
We now present the experimental evaluation of \SystemName{}.
Our evaluation answers the following questions:
\begin{enumerate}
    \item How does \SystemName{} perform in terms of wall-clock time, network costs and compute costs compared to state-of-the-art \ac{DL} systems (\Cref{sec:exp_comparison_fl})?
    \item What is the communication overhead of \SystemName{} (\Cref{sec:exp_overhead})?
    \item How does the sample size $ s $ and success fraction $ sf $ affect the performance of \SystemName{} (\Cref{sec:exp_parameters})?
    \item How effective is \SystemName{} at keeping views synchronized (\Cref{sec:exp_view_inconsistencies})?
\end{enumerate}

We implement \SystemName{} in the Python 3 programming language, spanning a total of \num{5488} lines of source code (SLOC).
\SystemName{} leverages the \texttt{IPv8} networking library~\cite{ipv8} which provides support for authenticated messaging and building decentralized overlay networks.
Our implementation adopts an event-driven programming model with the \texttt{asyncio} library.
We use the \texttt{PyTorch} library~\cite{paszke2019pytorch} to train ML models, and the dataset API from \textsc{DecentralizePy}~\cite{dhasade2023decentralized}.
As a node might be involved in multiple incoming and outgoing model transfers simultaneously, we equip each node with a bandwidth scheduler that we implemented.
This scheduler coordinates all model transfers a particular node is involved in.
Our implementation is open source and all artifacts are published in a GitHub repository.\footnote{See \url{https://github.com/devos50/decentralized-learning}.}

\subsection{Experiment Setup}
We run all experiments on machines in our national compute cluster.
Each machine is equipped with dual 24-core AMD EPYC-2 CPU, 128 GB of memory, an NVIDIA RTX A4000 GPU, and is running CentOS 8.
Similar to related work in the domain, we simulate the passing of time in our experiments~\cite{lai2022fedscale,lai2021oort,abdelmoniem2023refl}.
We achieve this by customizing the default event loop provided by the \texttt{asyncio} library and processing events without delay.
This requires minimal changes to the code, and therefore our implementation can be made suitable for usage in a production environment with trivial changes.

\textbf{Traces.}
We have designed \SystemName{} to operate in highly heterogeneous environments, such as mobile networks.
To verify that \SystemName{} also functions in such environments, we adopt various real-world traces to simulate pairwise network latency, bandwidth capacities, compute speed, and availability.
To model a WAN environment, we apply latency to outgoing network traffic at the application layer to realistically model delays in sending \SystemName{} control messages.
To this end, we collect ping times from WonderNetwork, providing estimations on the RTT between their servers located in 277 geo-separated cities~\cite{WonderNetwork}.
After data cleaning, we are left with a complete latency matrix for 227 cities worldwide.
When starting an experiment, we assign peers to each city in a round-robin fashion and delay outgoing network traffic accordingly.

We also adopt traces from the FedScale benchmark to simulate the hardware performance of nodes, specifically network and compute capacities~\cite{lai2022fedscale}.
These traces contain the hardware profile of 131'000 mobile devices and are originally sourced from the AI benchmark~\cite{ignatov2019ai} and the MobiPerf measurements~\cite{huang2011mobiperf}.
We assume that nodes are aware of the bandwidth capabilities of other nodes, and within a sample, a node sends its trained model to the aggregator with the highest bandwidth capability in the next sample.
Additionally, we use a trace containing the availability patterns of 136'000 mobile devices~\cite{yang2021characterizing}.
A device in these traces is considered online when it is charging and connected to a WiFi network (to minimize the impact on the user when training).
These availability traces show a diurnal pattern and reveal that most devices are only online for a few minutes~\cite{lai2022fedscale}.
Given the strictness of these availability traces where in the best scenario, only 8.8\% of devices are online, we assume graceful leaving.
In other words, when a node goes offline in \SystemName{}, ongoing model transfers and training will be interrupted immediately, but the node will notify $ 10 \cdot s $ nodes in its \view about leaving the system by sending a \texttt{membership} message.
In summary, our experiments go beyond existing work on \ac{DL} by integrating multiple traces that together account for the system heterogeneity and churn in WAN environments.

\begin{figure*}[t]
	\centering
     \begin{subfigure}{.9\textwidth}
    	\centering
    	\includegraphics[width=\linewidth]{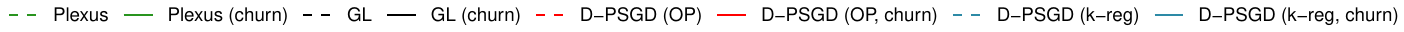}
    \end{subfigure}
	\begin{subfigure}{.5\columnwidth}
		\centering
		\includegraphics[width=\linewidth]{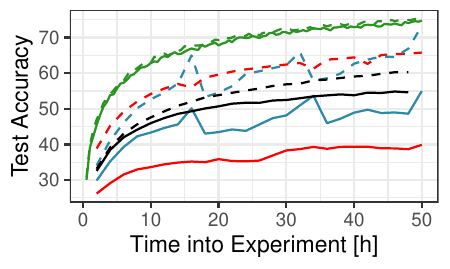}
		\label{fig:exp_accuracy_cifar10_iid}
	\end{subfigure}%
	\begin{subfigure}{.5\columnwidth}
		\centering
		\includegraphics[width=\columnwidth]{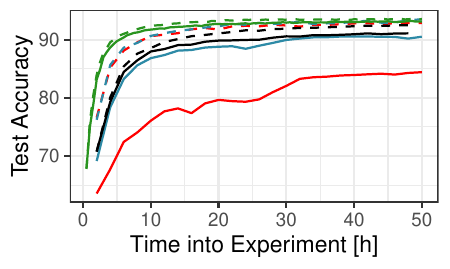}
		\label{fig:exp_accuracy_celeba}
	\end{subfigure}%
	\begin{subfigure}{.5\columnwidth}
		\centering
		\includegraphics[width=\columnwidth]{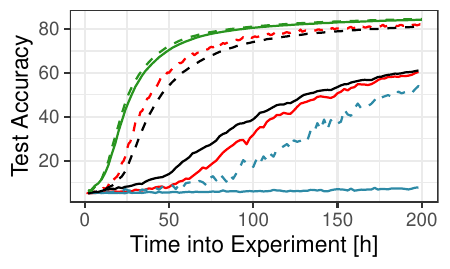}
		\label{fig:exp_accuracy_femnist}
	\end{subfigure}\vspace{-15pt}
	\begin{subfigure}{.5\columnwidth}
		\centering
		\includegraphics[width=\columnwidth]{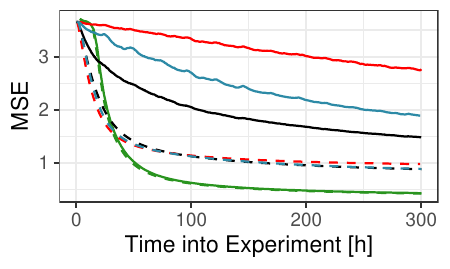}
		\label{fig:exp_accuracy_movielens}
	\end{subfigure}
	\begin{subfigure}{.5\columnwidth}
		\centering
		\includegraphics[width=\linewidth]{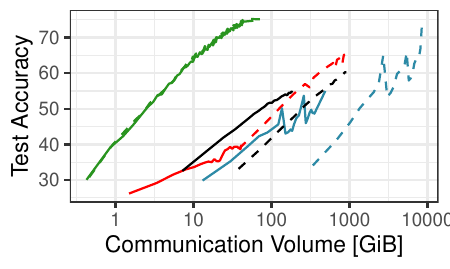}
		\label{fig:exp_communication_cifar10_iid}
	\end{subfigure}%
	\begin{subfigure}{.5\columnwidth}
		\centering
		\includegraphics[width=\columnwidth]{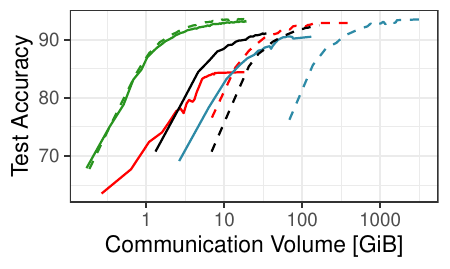}
		\label{fig:exp_communication_celeba}
	\end{subfigure}%
	\begin{subfigure}{.5\columnwidth}
		\centering
		\includegraphics[width=\columnwidth]{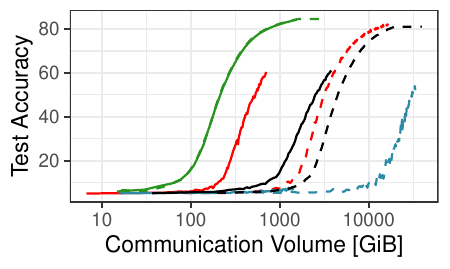}
		\label{fig:exp_communication_femnist}
	\end{subfigure}%
	\begin{subfigure}{.5\columnwidth}
		\centering
		\includegraphics[width=\columnwidth]{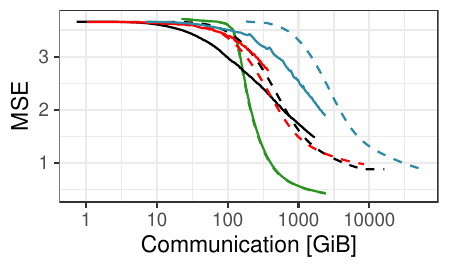}
		\label{fig:exp_communication_movielens}
	\end{subfigure}\vspace{-15pt}
	\begin{subfigure}{.5\columnwidth}
		\centering
		\includegraphics[width=\linewidth]{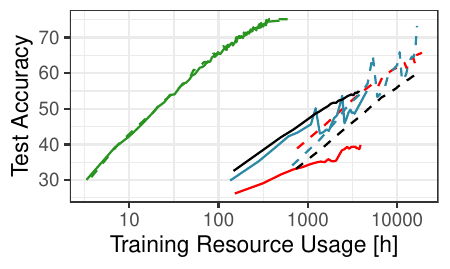}
		\caption{\cifar (IID)}
		\label{fig:exp_resource_cifar10_iid}
	\end{subfigure}%
	\begin{subfigure}{.5\columnwidth}
		\centering
		\includegraphics[width=\columnwidth]{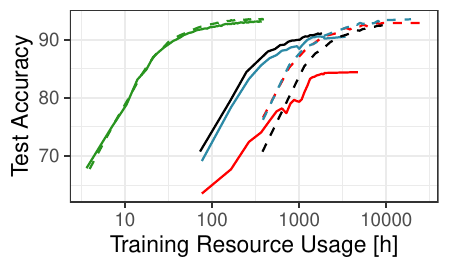}
		\caption{\celeba (non-IID)}
		\label{fig:exp_resource_celeba}
	\end{subfigure}%
	\begin{subfigure}{.5\columnwidth}
		\centering
		\includegraphics[width=\columnwidth]{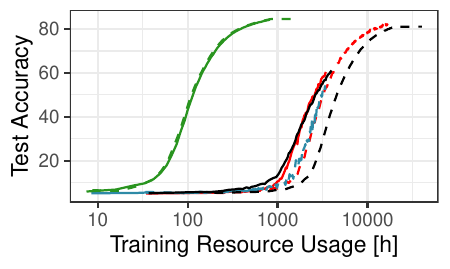}
		\caption{\femnist (non-IID)}
		\label{fig:exp_resource_femnist}
	\end{subfigure}%
	\begin{subfigure}{.5\columnwidth}
		\centering
		\includegraphics[width=\columnwidth]{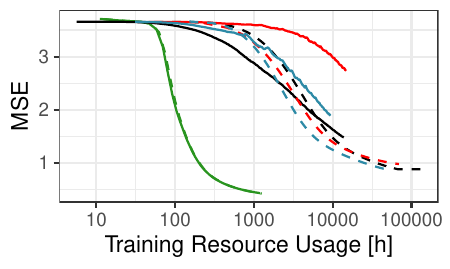}
		\caption{\movielens (non-IID)}
		\label{fig:exp_resource_movielens}
	\end{subfigure}
	\caption{The model convergence (top row), communication volume (middle row), and training resource usage (bottom row) for \SystemName{}, \ac{GL} and \dpsgd, for four learning tasks. We evaluate \dpsgd both with a one-peer (OP) exponential graph and a $k$-regular graph with $ k = 10 $.}
	\label{fig:exp_comparison_fl}
\end{figure*}

\textbf{Datasets.}
We evaluate \SystemName{} on different models and with four distinct datasets, whose characteristics are displayed in Table~\ref{table:experiment_datasets}.
The \cifar dataset~\cite{krizhevsky2009learning} is IID, partitioned by uniformly randomly assigning data samples to nodes. %
The \celeba and \femnist datasets are taken from the \leaf benchmark~\cite{caldas2018leaf}, which was specifically designed to evaluate the performance of learning tasks in non-IID settings.
Lastly, we also study a recommendation model based on matrix factorization~\cite{korenmatrixfactorization2009} on the \movielens 100K dataset~\cite{grouplens:2021:movielens}, comprising user ratings for several movies.
Here, we consider the one-user-one-node setup, imitating the realistic scenario where individual participants may wish to learn from the movie preferences of others.
The model sizes we use (\Cref{table:experiment_datasets}) range between \SI{124}{\kibi\byte} to \SI{6.7}{\mebi\byte}, in line with real-world deployments at scale on edge devices~\cite{bonawitz2019towards}.
Our evaluation, thus, covers a variety of learning tasks and data partitions.

\textbf{Performance metrics and hyperparameters.}
We measure the performance of the model on a global test set unseen during training, for the purpose of evaluation.
For the image classification tasks, we report the average top-1 test accuracy, while for the recommendation task, we report the mean squared error (MSE), indicating the difference between the actual and predicted ratings.
In line with other work, we fix the batch size to 20 for all experiments and each device always performs five local steps when training its model~\cite{lai2022fedscale,abdelmoniem2023refl} before communicating.
All models are trained using the SGD optimizer.
All our learning parameters (see~\Cref{table:experiment_datasets}) were adopted from previous works~\cite{caldas2018leaf, bellet2021d} or were considered after trials on several values. They yield acceptable target accuracy for all evaluated datasets.
We run each experiment three times with different seeds and report averaged values.

\subsection{\SystemName{} Compared to \ac{DL} Baselines}
\label{sec:exp_comparison_fl}
We now quantify and compare the performance of \SystemName{} with baseline \ac{DL} systems.

\begin{table*}[t]
	\small
	\centering
 \caption{A summary of experimental findings for the different datasets in the churn setting in Figure~\ref{fig:exp_comparison_fl}. For each dataset, we compare \SystemName{} against the best-performing baseline in terms of the highest individual model accuracy achieved across all nodes. We consider time-to-accuracy (TTA), communication-to-accuracy (CTA), and training-resources-to-accuracy (RTA).}
	\begin{tabular}{|c||c|c|c|c|c|c||c|c|c|}
		\hline
        \multirow{2}{*}{\textbf{Dataset}} & \textbf{Exp. Duration} & \multirow{2}{*}{\textbf{Method}} & \textbf{Best} & \textbf{TTA} & \textbf{CTA} & \textbf{RTA} & \multicolumn{3}{c|}{\cellcolor{lightgray!25} Savings by \SystemName{}} \\
		 & [\SI{}{\hour}] & & [\% or MSE] & [\SI{}{\hour}] & [\SI{}{\gibi\byte}] & [\SI{}{\hour}] & \cellcolor{lightgray!25} \textbf{TTA} & \cellcolor{lightgray!25} \textbf{CTA} & \cellcolor{lightgray!25} \textbf{RTA} \\ \hline \hline
		\multirow{2}{*}{\cifar} & \multirow{2}{*}{50} & \SystemName{} & 75.8 & 12.9 & 12.3 & 96.2 & \cellcolor{lightgray!25} & \cellcolor{lightgray!25} & \cellcolor{lightgray!25} \\
		 & & GL & 65.9 & 48 & 187.8 & 3840 & \multirow{-2}{*}{\cellcolor{lightgray!25}3.7$\times$} & \multirow{-2}{*}{\cellcolor{lightgray!25} 15.3$\times$} & \multirow{-2}{*}{\cellcolor{lightgray!25} 39.9$\times$} \\ \hline

		\multirow{2}{*}{\celeba} & \multirow{2}{*}{50} & \SystemName{} & 93.5 & 38.6 & 14.9 & 287 & \cellcolor{lightgray!25} & \cellcolor{lightgray!25} & \cellcolor{lightgray!25} \\
		 & & GL & 92.3 & 48 & 35.3 & 1844 & \multirow{-2}{*}{\cellcolor{lightgray!25} 1.2$\times$} & \multirow{-2}{*}{\cellcolor{lightgray!25} 2.4$\times$} & \multirow{-2}{*}{\cellcolor{lightgray!25} 6.4$\times$} \\ \hline

        \multirow{2}{*}{\femnist} & \multirow{2}{*}{200} & \SystemName{} & 84.5 & 38.3 & 308.1 & 163.5 & \cellcolor{lightgray!25} & \cellcolor{lightgray!25} & \cellcolor{lightgray!25} \\
		 & & GL & 64.2 & 198 & 3827.4 & 4353 & \multirow{-2}{*}{\cellcolor{lightgray!25} 5.2$\times$} & \multirow{-2}{*}{\cellcolor{lightgray!25} 12.4$\times$} & \multirow{-2}{*}{\cellcolor{lightgray!25} 26.6$\times$} \\ \hline

        \multirow{2}{*}{\movielens} & \multirow{2}{*}{300} & \SystemName{} & 0.44 & 36 & 291.6 & 140 & \cellcolor{lightgray!25} & \cellcolor{lightgray!25} & \cellcolor{lightgray!25} \\
		 & & GL & 1.44 & 300 & 1680 & 13302 & \multirow{-2}{*}{\cellcolor{lightgray!25} 8.3$\times$} & \multirow{-2}{*}{\cellcolor{lightgray!25} 5.8$\times$} & \multirow{-2}{*}{\cellcolor{lightgray!25} 370$\times$} \\ \hline
	\end{tabular}
	\label{table:exp_results}
\end{table*}

\textbf{Baselines and Setup.}
We use \Acf{GL}~\cite{hegedHus2019gossip} and \dpsgd as \ac{DL}~\cite{lian2017can} baselines.
In each round of \ac{GL}, a node first waits for some time and then sends its model to another random node in the network.
The selection of nodes is facilitated by a peer-sampling service which presents a view of random nodes in the network every round.
Upon receiving a model from another node, the recipient node merges it with its own local model, weighted by the model age, and trains the local model. 
\ac{GL} naturally tolerates churn and is robust to failing nodes.
However, pairwise model aggregation still leaves residual variance and deteriorates model convergence compared to when using global aggregation.
In our experiments, we fix the round timeout to 60 seconds for \ac{GL} to give each node sufficient time to train and transfer the model each local round.

\dpsgd~\cite{lian2017can} is a synchronous algorithm that only proceeds when all nodes have received all models from their neighboring nodes and therefore cannot be evaluated in churn settings without modifications.
We address this by using a fixed round duration for \dpsgd that we tune from the non-churn scenario by determining the amount of time it for 99\% of all model transfers to complete.
We specifically tolerate some model transfers to not complete as we found that otherwise round durations become disproportionally high because some network links have low bandwidth capacities.
We determine this timeout value separately for each dataset as they use models of differing sizes.
When the round timeout triggers, any pending model transfer is terminated.
We do not consider the cost of establishing the graph topology, which requires global coordination to spread edges exactly evenly between all nodes before the training starts.
We evaluate \dpsgd under two topologies: a 10-regular topology (\ie each node has ten neighbors) and a one-peer exponential graph topology, the latter being a state-of-the-art topology in \ac{DL}~\cite{ying2021exponential}.
Thus, we evaluate \dpsgd with both sparse and dense graph connectivity.

For \SystemName{}, we report the accuracy/MSE of the global model after aggregation every ten rounds.
For \dpsgd, we determine the mean and standard deviation of the accuracy obtained by evaluating models of individual nodes on the test dataset every two hours. 
However, for the \movielens dataset in \dpsgd, we report the MSE of the average model across nodes.
Since we use the one-user-one-node setup in the non-IID partitioning of the \movielens dataset, many users have missing embeddings for many movies which results in arbitrarily high losses on the global test if the prior approach of evaluating individual models is used.
We also report communication volume (transmitted bytes) and training resource usage (i.e., the time a device spends on model training).
For \SystemName{}, we set $\Delta t_{agg} = 300 $, $ s = 13 $ and $ sf = 0.8 $, i.e., an aggregator will complete aggregation when it receives at least 10 trained models or times out after 5 minutes, similar to~\cite{abdelmoniem2023refl}.
We run experiments with \cifar and \celeba for 50 hours, \femnist for 200 hours, and \movielens for 300 hours.

\textbf{Results.}
We show the performance of \SystemName{} and \ac{DL} baselines in~\Cref{fig:exp_comparison_fl}.
We evaluate the systems with (solid lines) and without (dashed lines) churn; in the latter setting all nodes always remain online.
The top row of~\Cref{fig:exp_comparison_fl} shows the test accuracy as the experiment progresses.
\SystemName{} outperforms both \ac{DL} baselines by converging quicker and achieving higher test accuracy, consistently across all datasets.
The performance of \SystemName{} is barely affected in the presence of difficult availability traces in the churn scenario, unlike \ac{GL} and \dpsgd.
In general, we find that in \ac{GL} more training occurs within a given time unit compared to \ac{DL}, since \ac{GL} rounds are asynchronous and individual nodes have less idle time compared to \dpsgd.
The performance of \dpsgd, in particular, is vulnerable to churn since a node will not exchange its model when a selected neighbor in the one-peer exponential graph is offline.
On the simpler binary classification task for the \celeba dataset, the performance improvements of \SystemName{} are modest.
However, on more difficult learning tasks like the 62-class image classification in \femnist with a larger model size, \SystemName{} achieves more than 20\% better accuracy when compared to the best performing \ac{DL} baseline, \ac{GL}.
The middle row in~\Cref{fig:exp_comparison_fl} shows the communication volume (horizontal axis, log scale) required to achieve the test accuracy for the evaluated systems.
\SystemName{} attains high test accuracies with orders of magnitude less transmitted bytes.
We note that \dpsgd with a 10-regular topology incurs the most network traffic while performing on par or worse than the one-peer exponential graph topology.
The bottom row in~\Cref{fig:exp_comparison_fl} shows the training resource usage (horizontal axis, log scale) consumed to achieve the test accuracy.
\SystemName{} attains high test accuracies with orders of magnitude less resource usage.

To better understand the improvements of \SystemName{} with the baselines, we summarize key numerical results from~\cref{fig:exp_comparison_fl} in~\Cref{table:exp_results}.
For each dataset, we determine the best-performing baseline in terms of the highest \emph{individual model accuracy achieved across all nodes}.
We remark that the accuracies in \Cref{fig:exp_comparison_fl} are different from values reported in \Cref{table:exp_results} as the former shows averaged accuracies.
We found that \ac{GL} for all datasets produced the model with the highest accuracy across all baselines.
We then determine time-to-accuracy (TTA), communication-to-accuracy (CTA), and resources-to-accuracy (RTA), which are metrics that represent the efficacy, efficiency, and scalability of \ac{DL} systems.
For the evaluated datasets and compared to the target accuracy, \emph{\SystemName{} saves 1.2$\times$ - 8.3$\times$ in TTA, 2.4$\times$ - 15.3$\times$ in CTA and 6.4$\times$ - 370$\times$ in RTA compared to \ac{GL} and \dpsgd}.
In conclusion, our comprehensive evaluation demonstrates the superior efficiency and effectiveness of \SystemName{}. %

\begin{figure}[b]
	\centering
     \begin{subfigure}{.8\columnwidth}
    	\centering
    	\includegraphics[width=\linewidth]{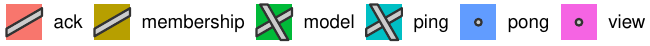}
    \end{subfigure}
	\begin{subfigure}{.5\columnwidth}
		\centering
		\includegraphics[width=\linewidth]{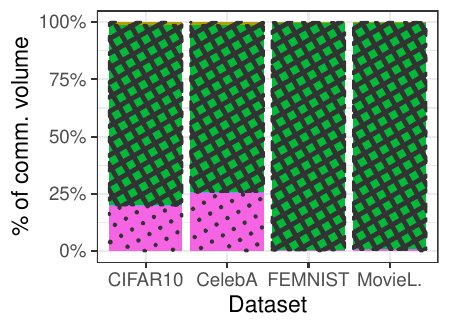}
        \caption{\% of total comm. volume.}
		\label{fig:modest_overhead_bytes}
	\end{subfigure}%
	\begin{subfigure}{.5\columnwidth}
		\centering
		\includegraphics[width=\columnwidth]{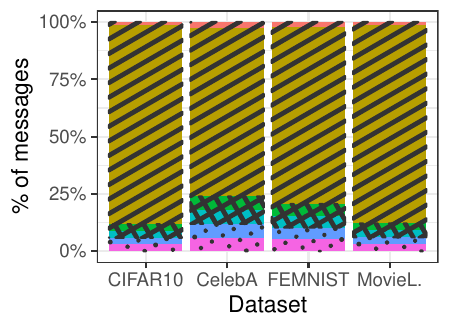}
        \caption{\% of messages.}
		\label{fig:modest_overhead_count}
	\end{subfigure}%
	\caption{Breakdown of network usage by \SystemName{} in the churn scenario, per message type and for each dataset.}
	\label{fig:modest_overhead}
\end{figure}

\subsection{Overhead of \SystemName{}}
\label{sec:exp_overhead}
The overhead of \SystemName{}, compared to other \ac{DL} algorithms, mainly comes from the additional network activity in \SystemName{}, e.g., \texttt{ping} and \texttt{pong} messages, and the exchange of \views across samples.
To quantify the network overhead of \SystemName{}, we show the distribution of communication volume (transmitted bytes) and the number of messages for each message type in~\Cref{fig:modest_overhead} during the experiments with churn~(\Cref{sec:exp_comparison_fl}).
\Cref{fig:modest_overhead_bytes} shows that across all datasets, most of the transmitted bytes comprise model transfers: 73.6\% to 99.2\% of all transmitted bytes for \celeba and \movielens, respectively, are model transfers.
Regarding \SystemName{}-exclusive messages, most overhead in bytes comes from \texttt{view} messages.
The size of a \texttt{view} message grows proportionally to the network size.
With $ n = 355 $, a \texttt{view} message is 31.3 \SI{}{\kibi\byte} in size which increases to 88.0 \SI{}{\kibi\byte} with $ n = \num{1000} $.
The magnitude of network traffic related to \SystemName{} also depends on the sample size, \textit{e.g.}, a larger sample size results in more \textsc{ping}, \textsc{pong} and \textsc{view} messages being sent.
\Cref{fig:modest_overhead_bytes} highlights that the network overhead of \SystemName{}, i.e., all messages except \texttt{model} messages, decreasing as the model size increases.
The overhead of \SystemName{} is minimal for the FEMNIST and MovieLens datasets, e.g., \SystemName{} only \emph{increases network traffic by 0.5\%} for FEMNIST.

In~\Cref{fig:modest_overhead_count} we show the absolute number of messages, as a fraction of the total number of messages transmitted.
For all datasets, the \texttt{membership} message is most commonly sent, originating from the churn in the availability traces.
We remark that each \texttt{model} message is accompanied by a \texttt{view} message to maintain view synchronization, and therefore they are sent in equal amounts.
The overhead of \texttt{membership} messages is indiscernible in~\Cref{fig:modest_overhead_bytes} since these messages are only 195 bytes in size.

Furthermore, \SystemName{} deliberately overutilizes the computational resources in order to deal with churn.
With $ sf = 0.8 $, at least 20\% of trained models will never be aggregated as the aggregator proceeds once it receives sufficient models or times out.
Therefore, work done by some of the participants will not be integrated into the aggregated model.
This may be termed the computational overhead of \SystemName{}.
Various learning systems alleviate this issue by integrating stale model updates~\cite{wu2020safa,damaskinos2022fleet,abdelmoniem2023refl}.
We consider the integration of stale models with \SystemName{} beyond the scope of this work.

\begin{figure}[b]
	\centering
     \begin{subfigure}{.5\columnwidth}
		\centering
		\includegraphics[width=\linewidth]{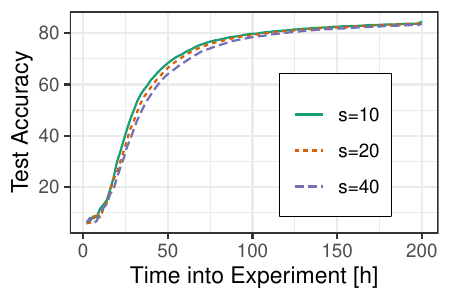}
        \caption{Test Accuracy}
		\label{fig:vary_s_accuracies}
	\end{subfigure}%
    \begin{subfigure}{.5\columnwidth}
		\centering
		\includegraphics[width=\linewidth]{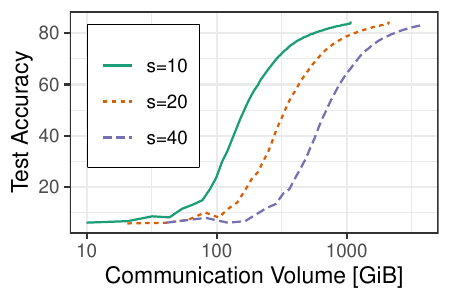}
        \caption{Communication Volume}
		\label{fig:vary_s_communication}
	\end{subfigure}
    \begin{subfigure}{.5\columnwidth}
		\centering
		\includegraphics[width=\linewidth]{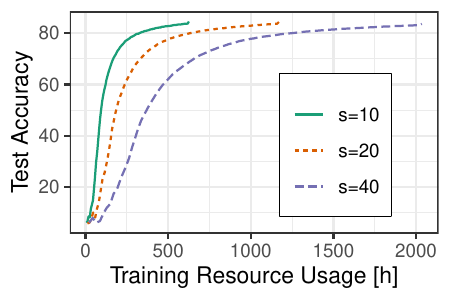}
        \caption{Training Resource Usage}
		\label{fig:vary_s_resource}
	\end{subfigure}%
    \begin{subfigure}{.5\columnwidth}
		\centering
		\includegraphics[width=\linewidth]{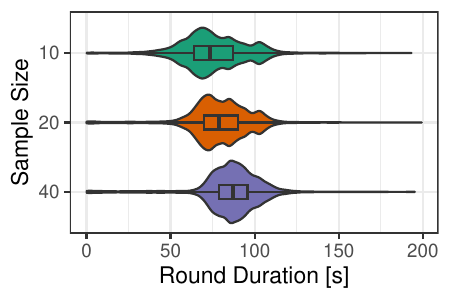}
        \caption{Round Duration}
		\label{fig:vary_s_round_durations}
	\end{subfigure}%
 
	\caption{The performance of \SystemName{} on the \femnist learning task, for different sample sizes $ s $.}
	\label{fig:vary_s}
\end{figure}

\subsection{The Effect of $ s $ and $ sf $ on \SystemName{} Performance}
\label{sec:exp_parameters}
\SystemName{} uses two parameters, namely sample-size ($s$) and success-fraction ($sf$) that dictate how the system performs.
We now quantify the effect of varying these parameters independently, on the performance of \SystemName{} in the presence of churn.
The following experiments use the \femnist dataset which has the largest model size in our setup.

\textbf{The effect of $ s $.}
We first explore the effect of $ s $ sample sizes on model convergence, communication volume, resource usage, and round duration by running \SystemName{} with $s = 10$, $20$, and $40$.
We show the results of this experiment, keeping $sf = 0.8 $ constant in~\Cref{fig:vary_s}.
\Cref{fig:vary_s_accuracies} shows for different values of $ s $ the test accuracy as the experiment progresses.
Around 50 hours into the experiment we observe that increasing $ s $ actually slows down training, likely because more models have to be transferred to and from the aggregator.
We also find that the achieved test accuracy after 200 hours of training with \SystemName{} is lower for $ s = 40 $ compared to $ s = 10 $: 84.6\% vs. 83.6\%.
Naturally, increasing $ s $ also has a negative impact on communication cost and resource usage, which are visualized in~\Cref{fig:vary_s_communication} and~\Cref{fig:vary_s_resource}, respectively.
To reach 83\% test accuracy with $ s = 40 $, \SystemName{} incurs 4.1$\times$ additional communication volume and 4.9$\times$ more training resource usage, when compared to $ s = 10 $.

Increasing $ s $ also prolongs the duration of individual rounds.
We show in~\Cref{fig:vary_s_round_durations} the distribution of round durations in seconds for different values of $ s $ using a box and violin plot.
For presentation clarity, we removed outlier round durations larger than \SI{200}{seconds}, e.g., when an aggregator triggered the aggregation timeout.
When increasing $ s $ from 10 to 40, the average round duration increases from \SI{75.7}{seconds}. to \SI{86.2}{seconds}.
At the same time, we observe also a positive effect on round duration when increasing $ s $: with higher values of $ s $, there is a higher probability that nodes with high bandwidth capacities are included in the sample compared to lower values of $ s $, which lowers the overall model transfer times during a round.
We can see this effect in~\Cref{fig:vary_s_round_durations} as there is a higher variance in round durations for lower values of $ s $.
Empirically, we obtain a good trade-off between sample size and convergence when setting the sample size around 10.

\begin{figure}[t]
	\centering
     \begin{subfigure}{.5\columnwidth}
		\centering
		\includegraphics[width=\linewidth]{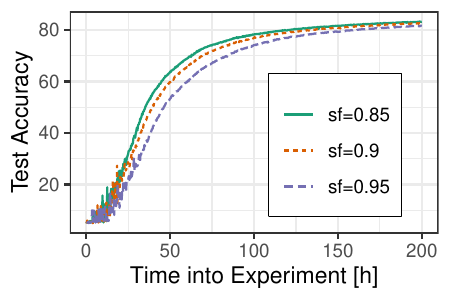}
        \caption{Test Accuracy}
		\label{fig:vary_sf_accuracies}
	\end{subfigure}%
    \begin{subfigure}{.5\columnwidth}
		\centering
		\includegraphics[width=\linewidth]{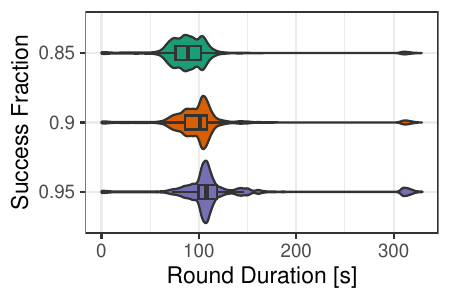}
        \caption{Round Duration}
		\label{fig:vary_sf_round duration}
	\end{subfigure}
	\caption{The performance of \SystemName{} on the FEMNIST learning task, for different success fractions $ sf $.}
	\label{fig:vary_sf}
\end{figure}

\textbf{The effect of $ sf $.}
Next, we show the effect of the success fraction on the performance \SystemName{}, see \Cref{fig:vary_sf}.
For this experiment, we fix $ s = 20 $ and consider $ sf = 0.85 $, $ sf = 0.9 $ and $ sf = 0.95 $.
Increasing the success fraction reduces the %
model updates that do not make it to the aggregation but prolongs the round duration as additional models have to be transferred before the next round can start.
Furthermore, increasing $ sf $ makes it more likely that an aggregator will not receive sufficient models to wrap up aggregation due to nodes going offline, and hence, trigger the aggregation timeout.
\Cref{fig:vary_sf_accuracies} shows the test accuracy for different values of $ sf $ as the experiment progresses, and highlights that increasing $ sf $ has a negative impact on model convergence.
To further inspect this, we also plot the distribution of round durations for the different values of $ sf $ in~\Cref{fig:vary_sf_round duration}.
We observe a 31.8\% increase in average round duration when increasing $ sf $ from 0.85 to 0.95: from \SI{92.3}{seconds} to \SI{121.7}{seconds}.
This is both because with larger values of $ sf $ an aggregator needs to send and receive more models, and because we observe more aggregation time-outs ($\Delta t_{agg} = 300 $ in our experiments).

\begin{figure}[t]
	\centering
     \begin{subfigure}{.7\columnwidth}
    	\centering
    	\includegraphics[width=\linewidth]{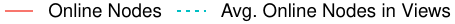}
    \end{subfigure}
	\begin{subfigure}{.5\columnwidth}
		\centering
		\includegraphics[width=\linewidth]{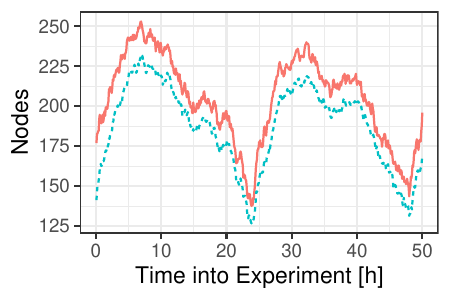}
        \caption{\cifar ($n = \num{1000} $)}
		\label{fig:exp_view_consistency_cifar}
	\end{subfigure}%
	\begin{subfigure}{.5\columnwidth}
		\centering
		\includegraphics[width=\columnwidth]{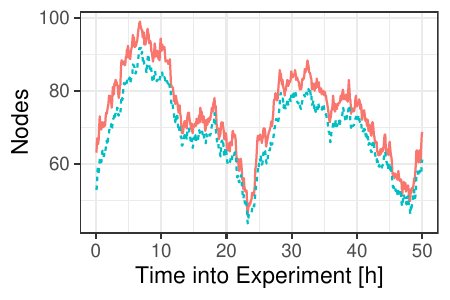}
        \caption{\femnist ($ n = 355 $)}
		\label{fig:exp_view_consistency_femnist}
	\end{subfigure}%
	\caption{The number of online nodes, and the average number of online nodes in the views of participating nodes for the \cifar and \femnist datasets.}
	\label{fig:exp_view_consistency}
\end{figure}

\subsection{View Inconsistencies in \SystemName{}}
\label{sec:exp_view_inconsistencies}
In \SystemName{}, nodes can leave and join the system at any time.
Therefore, \views and propagation of changes in these views across the network are critical to the functioning of \SystemName{}.
Here, we evaluate the ability of \SystemName{} to propagate changes in \views. %
Specifically, we focus on how the views of different nodes are updated as the experiment progresses.
For this experiment, we quantify the \emph{actual number of online nodes} and compare it against the \emph{average number of online nodes in the views of online nodes} every five minutes over a run of \SystemName{}. 
We omit the views of offline nodes as they are likely to have outdated views.
As we suspect more inconsistencies among the \views in large networks, we run this experiment both for \cifar (with $ n = \num{1000} $) and \femnist (with $ n = 355 $) for a duration of \SI{50}{\hour}.

\Cref{fig:exp_view_consistency} shows the actual number of online nodes (red) and the average number of online nodes across the views (blue) at any given time.
We observe %
a diurnal pattern (as seen in FedScale~\cite{lai2022fedscale} and REFL~\cite{abdelmoniem2023refl}) with the number of online nodes peaking around 7 hours into the experiment.
The difference between the two lines indicates the level of synchronization of views between nodes.
In general, we note that the trend of the number of online nodes in the \views is consistent with the same trend as the actual number of online nodes.
Furthermore, we can observe that the views of \SystemName{} nodes usually contain fewer online nodes than there are actually online.
This is a direct consequence of the combined effects of (1) latency of forwarding the membership of a newly joining node to \emph{all nodes}, and (2) other nodes going offline during this propagation.
This effect is more pronounced in \Cref{fig:exp_view_consistency_cifar} than in~\Cref{fig:exp_view_consistency_femnist}, which is because the network size for \cifar ($ n = \num{1000} $ is larger than for \femnist ($ n = \num{355} $).
In particular, it is more challenging to keep up-to-date views when the network size grows.
This can be offset by either increasing the sample size (such that membership information spreads to more nodes during a round) or by disseminating \texttt{advertise} messages to more nodes.

To conclude, \SystemName{} maintains a high degree of synchronization of \views.
In addition, we have not noticed any issues arising from these weakly consistent views in our experiments, further strengthening our claim.

\section{Related Work}
\label{sec:related}

\textbf{Decentralized Learning (DL).}
\dpsgd~\cite{lian2017can}, also known as D-SGD, showed theoretically and empirically that under strong bandwidth limitations on an aggregation server in a data center, decentralized algorithms can converge faster. %
Assumptions on the behavior of those algorithms make them most suited to data centers.
The synchronization required in \dpsgd is costly when training on edge devices.
As a solution, research in \ac{DL} has been focussed either on having a better topology~\cite{bellet2021d, vogels2022beyond, ying2021exponential, song2022communicationefficient},
or designing asynchronous algorithms~\cite{pmlr-v80-lian18a, ormandi2013gossip}.
MoshpitSGD~\cite{ryabinin2021moshpit} uses a DHT to randomly combine nodes in multiple disjoint cohorts every round for fast-averaging convergence.
All these algorithms rely on all nodes being online in each round, while also overlooking system heterogeneity.
On the other hand, HADFL is a framework that conducts fully asynchronous training on heterogeneous devices~\cite{cao2021hadfl}.
HADFL probabilistically selects neighboring nodes to exchange the model with but relies on a central server for coordination, unlike \SystemName{}.
In contrast to these algorithms, \SystemName{} has been designed and optimized for the edge scenario where nodes can leave and join at any time, without relying on any central coordination.

\textbf{Federated Learning (FL).}
Federated Learning is arguably the most popular algorithm for privacy-preserving distributed learning and uses a parameter server to coordinate the learning process~\cite{pmlr-v54-mcmahan17a}.
Similar to \SystemName{}, FL lowers resource and communication costs at the edge by having a small subset of nodes train the model every round.
To make FL suitable at scale in deployment scenarios, recent works have placed significant emphasis on system challenges~\cite{bonawitz2019towards,MLSYS2022_f340f1b1,pmlr-v151-nguyen22b,lai2021oort,abdelmoniem2023refl}.
FL still requires a highly-available central server that can support many clients concurrently, possibly resulting in high infrastructure costs.
\SystemName{}, on the contrary, is a fully decentralized system with an aggregation scheme inspired by \ac{FL}, while avoiding central coordination.

\textbf{Blockchain-Assisted \ac{DL}.}
We identified various works that implement and evaluate blockchain-based decentralized learning systems~\cite{lu2019blockchain,pokhrel2020federated,majeed2019flchain,bao2019flchain}.
Consensus-based replicated ledgers used in these systems provide strong consensus primitives at a significant and unnecessary overhead.
Machine learning optimizations based on SGD thrive in the presence of stochasticity, obviating the need for strong consensus in the form of a global model~\cite{lian2017d-psgd,ormandi2013gossip}.

\textbf{Decentralized Peer Sampling.} Brahms~\cite{bortnikov2009brahms}, Basalt~\cite{auvolat2021scriptstyle} and PeerSampling~\cite{jelasity2007gossip} provide uniformly random samples without network-wide synchronization.
However, to the best of our knowledge, \SystemName{} is the first decentralized sampling mechanism that ensures samples are mostly consistent, in the presence of nodes joining and leaving.

\section{Conclusions}
\label{sec:conclusion}
This paper introduced \SystemName{}, a practical and efficient \ac{DL} system for training in cross-device edge settings.
The two key components of our system are a decentralized peer sampler to select small subsets of nodes each round, and a global aggregation operation within these selected subsets.
A cardinal property of \SystemName{} is the ability to deal with churn, i.e., nodes joining and leaving.
Extensive evaluations with realistic traces of compute speed, network capacity, and availability in decentralized networks up to \num{1000} nodes demonstrate the superiority of \SystemName{} over baseline \ac{DL} algorithms, reducing time-to-accuracy by 1.2-8.3$\times$, communication volume by 2.4-15.3$\times$, and training resource usage by 6.4-370$\times$.
Overall, \SystemName{} represents a significant advancement in practical decentralized learning systems, paving the way for deployments in cross-device edge networks without any central server.

\bibliography{references}{}
\bibliographystyle{IEEEtran}

\end{document}